\catcode`\@11
\documentclass[11pt]{article}
\usepackage{dsfont}
\usepackage{times}\usepackage{cite} \usepackage{amssymb}
\usepackage{pifont}\usepackage{amsmath}\usepackage{amscd}
\usepackage{eepic}\usepackage{psfig}\usepackage{subfigure}
\usepackage[all]{xy}
\numberwithin{equation}{section}
\numberwithin{table}{section}
\newfont\wasy{wasy10 scaled 2000}

\newfont\calig{pzcmi}

\newcommand\remove[1]{}
\newcommand\preprint[1]{\vspace{-1in}\vtop{\null\hfill
\parbox[t]{1.6in}{\small\sc #1\\\null}}
\vskip .5in\bigskip\normalfont}
\newdimen\arrayruleHwidth
\makeatletter
\def\Hline{\noalign{\ifnum0=`}\fi\hrule \@height
\arrayruleHwidth \futurelet
\@tempa\@xhline}
\makeatother

\renewcommand\bar\overline
\newcommand\cf[4]{\bibitem{#1}{#2}.~{\it #3};~{#4}.}

\def\C{{\mathds{C}}}

\newcommand\dual[1]{\tilde{#1}}
\newcommand\End[1]{\operatorname{End} {#1}}
\newcommand\Ext[1]{\operatorname{Ext} ({#1})}

\def\eq#1{(\ref{#1})}

\def\goth#1{{\mathfrak #1}}
\renewcommand\hat{\widehat}
\newcommand\Hom[1]{\operatorname{Hom}{(#1})}

\newcommand\ie{{\slshape i.e.~}}

\def\pa{\partial}

\def\rt{\longrightarrow}

\def\tilde{\widetilde}

\def\Z{\mathds{Z}}
\newfont\sheafnt{rsfs10}

\newcommand\atmp[3]{Adv. Theor. Math. Phys. {\bf #1}~(#2)~#3}

\newcommand\hepth[1]{arXiv:hep-th/{#1}}

\newcommand\jhep[3]{JHEP {\bf {#1}}~({#2})~{#3}}

\newcommand\npb[3]{Nucl.~Phys. {\bf B{#1}}~({#2})~{#3}}

\newcommand\prd[3]{Phys. Rev. {\bf D{#1}}~({#2})~{#3}}

\newcommand\rmp[3]{Rev. Mod. Phys. {#1}~({#2})~{#3}}
\newcommand\fortsch[3]{Fortsch. Phys. {#1}~({#2})~{#3}}
\begin{document}
\title{\preprint{hep-th/0309191}
Seiberg duality as derived equivalence for some quiver gauge
theories}
\author{ \parbox[h]{.4\textwidth}{Subir Mukhopadhyay
\thanks{subir@physics.umass.edu} \\ 
\calig Department of Physics \\
\calig University of Massachusetts \\
\calig Amherst, MA 01003-4525.  USA}
\and 
\parbox[h]{.4\textwidth}{
\raggedright
Koushik Ray \thanks{koushik@iacs.res.in} \\
\calig Department of Theoretical Physics\\
\calig Indian Association for  the Cultivation of Science\\
\calig 2A \& B Raja S~C~Mullick Road,
\calig Calcutta 700~032. India}}
\date{}
\maketitle
\begin{abstract}
\noindent We study Seiberg duality of quiver gauge theories 
associated to the complex cone over the second del Pezzo surface.
Homomorphisms in the path algebra of the quivers in each of 
these cases satisfy  relations which follow from a superpotential 
of the corresponding gauge theory as F-flatness conditions.  
We verify that Seiberg 
duality between each pair of these theories can be understood 
as a derived equivalence between the categories
of modules of representation of the path algebras of the 
quivers. Starting from the projective
modules of one quiver we construct tilting complexes  whose 
endomorphism algebra yields the path algebra of the dual 
quiver. Finally, we present a general scheme for obtaining
Seiberg dual quiver theories by constructing quivers whose
path algebras are derived equivalent.
We also discuss some combinatorial relations between 
this approach and some of the  other approaches which has been used to
study such dualities.
\end{abstract}
\vspace{.7cm}



\vspace{.7cm}

\thispagestyle{empty}
\newpage

\tableofcontents
\section{Introduction}
Seiberg duality relates two $\mathsf{N} = 1 $ supersymmetric gauge 
theories with product gauge groups
\cite{seiberg,brodie,kutasov,intril}. 
Two Seiberg dual theories have gauge groups with
factors of different ranks $N_c$, but the same number of
flavors $N_f$. 
A pair of theories which are Seiberg dual to each other 
belong to the same universality 
class; that is, they flow to the same conformal theory in the 
infra-red (IR).
A physical way to realise this duality is by means of the suspended 
brane configuration \cite{hanany,diaconescu,elitzur}. In this 
scheme first a gauge theory is realised as the world volume theory 
of D-branes suspended between an
$\mathrm{NS}5$ and an $\mathrm{NS}5'$ brane. Seiberg duality is then 
interpreted as a transporting the $\mathrm{NS}5$ brane through the 
adjacent $\mathrm{NS}5'$ brane, by invoking the so called s-rules.

An interesting class of theories where explicit
constructions and tests of Seiberg duality have already  been 
carried out successfully consists of quiver gauge theories
. 
Some quiver gauge theories \cite{dm} can be obtained 
as world-volume theories of D-branes on an
orbifold. Seiberg duality of a pair of such theories relates 
the corresponding D-branes, which in turn correlates this 
field-theoretic duality to the geometry of the moduli space 
of the gauge theory. Indeed, two Seiberg dual
quiver gauge theories of D-branes turn out to have the same 
vacuum moduli space. In the special case in which the rank of the 
gauge group on the \emph{dualising node}, the one 
selected to perform Seiberg duality, remains
unchanged, Seiberg duality is amenable to a toric description.
This happens if the rank of the gauge group associated to
this node, say $i_0$, and the corresponding flavors are
related as $N_f(i_0) = 2 N_c(i_0)$.
The toric description derives from the interpretation 
of the vacuum moduli space of the associated  gauge 
theories of D-branes as toric varieties. 
In such cases, where both theories of the dual couple
have a toric moduli space, the actions of the algebraic
torus $\C^{\star}$ on the toric varieties restrict the 
pair of gauge groups to be $U(1)$ in the
abelian and $SU(N)$ in the non-abelian case, with the same $N$. 
This special case of Seiberg duality is known as 
\emph{toric duality} \cite{fhh1}. Many examples of 
Seiberg dual pairs of D-brane gauge theories have been 
constructed using toric duality 
\cite{beasley,fhh1,fhh2,fhhu,ffhh,fh}.
The basic idea underlying toric duality is to construct a quiver gauge 
theory as the world-volume gauge theory of D-branes on the
three-dimensional orbifold $\C^3/\Z_3\times\Z_3$ \cite{fhhu,tapo}. 
Then by considering different partial
resolutions of this orbifold one manufactures different
gauge theories 
with the same IR limit, namely the theory at the orbifold.
The complex cones over the zeroth 
Hirzebruch surface and the second del
Pezzo surface arising from such partial resolutions each
gives rise 
to a pair of toric dual theories \cite{fhh1,fhh2,morples}. The complex 
cone over the third del Pezzo surface, arising from another set 
of  partial resolutions of $\C^3/\Z_3\times\Z_3$ gives
rise to four toric dual theories instead\cite{fhh2,beasley}. We shall 
summarise the rules for obtaining toric dual theories in the 
next section.

There is another approach to Seiberg duality, 
however \cite{berdoug}. The main idea behind this approach 
is to view  a pair of Seiberg dual
theories as the very same gauge theory, but
constructed in terms of a different set of basic D-branes.
The category of
D-branes in $\mathsf{N}=1$ supersymmetric quiver gauge theories, 
many of which are realised on an orbifold,
is the bounded derived category 
of representations of the the path-algebra of the quiver. 
In this point of view
it is  possible to part company with the
orbifolds and consider manifestations of Seiberg duality in this 
category. It turns out that Seiberg duality is a derived functor 
in this category --- a pair of Seiberg dual theories are derived 
equivalent. This approach uses Rickard's theorem \cite{rickard} 
on derived equivalence of the bounded derived category  of
representations of algebras.  Let 
$D^b({\cal A}\operatorname{-mod})$ denote the bounded derived 
category of representations of an algebra $\cal A$ over the
field of complex numbers, $\C$ and 
let $T$ be a complex in $D^b({\cal A}\operatorname{-mod})$. The 
complex $T$ is called a \emph{tilting complex} if the algebra 
$\Hom{T,T}$ has finite global dimension, the direct summands of 
copies of $T$ generate $D^b({\cal A}\operatorname{-mod})$ and 
its self-extension vanishes, $\Ext{T,T}=0$. Then, if the 
endomorphism algebra $\End{T}$ of a 
tilting complex $T$ in $D^b({\cal A}\operatorname{-mod})$ coincides 
with an algebra $\cal B$, the algebras $\cal A$ and $\cal B$ 
are derived equivalent. Applying this to the  path algebras of 
quivers, if ${\cal A} = \C Q$, the path algebra of a quiver $Q$ 
over $\C$ and ${\cal B} = \C Q'$, then these two path algebras are 
derived equivalent if a tilting complex $T$ can be
constructed in 
$D^b({\C Q}\operatorname{-mod})$ whose endomorphism algebra coincides 
with $D^b({\C Q'}\operatorname{-mod})$ and vice-versa. The
quivers $Q$ and $Q'$ will be termed \emph{dual} in the sequel and
we shall use the shorthand $D(Q)$ for $D^b({\C
Q}\operatorname{-mod})$.
It has been conjectured \cite{berdoug} that
two gauge theories corresponding to quivers whose path algebras
are derived equivalent in this sense are Seiberg dual to each other. 
Let us stress at this point that this way of obtaining Seiberg dual 
theories does not depend on the toric description of the moduli 
space of the gauge theories. Thus, it is possible to deal with 
theories with $N_f \neq 2N_c$. Furthermore, in a physically interesting 
gauge theory one has, apart from the ``quarks" spanning the path algebra 
of the quiver, relations which follow from a
superpotential as F-flatness conditions. 
It has been explicitly checked in certain examples \cite{braun} that 
the dual fields in the gauge theory corresponding to the dual quiver 
constructed through derived equivalence satisfy relations that follow 
from the F-flatness conditions of the \emph{dual
superpotential}, that is the superpotential of the dual
gauge theory. The dual superpotential is a deformation of the
original superpotential one start with, by relevant
operators, expressed in terms of dual fields, as we describe
below. What is more, the dual superpotential has but cubic
interactions, thereby guaranteeing $\mathsf{N}=1$
supersymmetry of the dual gauge theory. These 
examples include the quivers arising from the complex cone over 
the zeroth Hirzebruch surface studied earlier \cite{fhh1,fhh2}.

In this note we verify this conjecture for the quivers
associated to the
second del Pezzo surface. There are only two toric 
phases for this case \cite{fhh2}. Hence there are only two
toric dual theories.
However, once we relax the restriction on the theory 
to be toric, \ie allow for $N_f\neq 2N_c$
on the dualising node, we show by explicit construction that 
there are other
theories, which are Seiberg dual to each other. For all these 
cases we construct
tilting complexes starting from one and find out the quivers
whose path algebra coincides with the endomorphism algebra 
of the tilting complex, that is, the dual quivers. 
We verify that the associated gauge theories are Seiberg
dual to each other and are compatible with the
known rules laid down earlier \cite{fhhk}. These non-toric 
theories are of
import in studying renormalisation group flow and the
resulting cascade of Seiberg dual theories
\cite{fiol,fhhk,hw}. The toric as well as the non-toric
phases arise along such a flow in general.

The lay out of this note is as follows.
In the next section we set up notations and recall
the rules for performing Seiberg duality operations on
a quiver in general. In \S\ref{delpez2} we discuss
Seiberg duality of the quivers associated with the
second del Pezzo surface. We explicitly construct tilting
complexes and derive the endomorphism algebras by taking
each of the five nodes of the starting quiver in turn. 
In \S\ref{discus} we present a general method for obtaining
Seiberg dual theories based on the examples of
\S\ref{delpez2}.
We summarise the results in \S\ref{concl}.
\section{Seiberg duality of quiver gauge theories}
In this note we restrict our discussions to $\mathsf{N}=1$ 
quiver gauge theories,
which have gauge groups of the form ${G} = \prod_i SU(N_c(i))$ 
and flavors
$N_f(i)$ associated to each node $i$ of the quiver. 
A quiver gauge theory contains chiral multiplets, which we 
shall refer to as \emph{quarks} at times. 
While each factor in the gauge group $G$ corresponds to a 
node in the quiver,  each
arrow corresponds to a chiral field transforming in the 
bi-fundamental representation of the two factors on the two 
nodes it connects.  We use the convention that a chiral field 
$\chi$ transforming in the
fundamental representation of a factor $SU(N_c(i))$ and 
in the anti-fundamental of
the (possibly same) factor $SU(N_c(j))$ in $G$ will be 
represented by an arrow 
from the $j$-th node to the $i$-th node of the quiver, and will be
denoted $\chi_{ij}$, as~
\entrymodifiers={+<.5em>[o][F-]}
$\xymatrix{j\ar[r]_{\hbox{$\chi_{ij}$}} & i}$. \entrymodifiers={}
For the morphisms between projectives, we use the convention \cite{braun} that
morphisms are along the arrows of the quiver. 

Seiberg duality provides a prescription to relate the chiral 
operators of two
theories. This can thus be used to map the deformations of one 
theory to that of
the other in such a way that the low-energy (IR) properties 
of the pair
remain unaltered. 
Let us briefly recall how this works \cite{beasley}. Starting 
with a quiver gauge
theory one chooses a factor in the gauge group $G$, say 
$SU(N_c(i_0))$,
corresponding to a node $i_0$ in the quiver. The chiral fields 
transforming in
the fundamental corresponding to this factor and represented 
by incoming arrows
on the node $i_0$ are singled out. The theory is then 
\emph{deformed} 
by turning off the superpotential couplings of these fields.
The gauge couplings of
all the factors $SU(N_c(i))$, $i\neq i_0$, in $G$ under which 
these fields
are charged are sent to infinity as well, \ie $1/g_i^2\rt 0$.
Seiberg duality then predicts an equivalent description of this 
deformed theory 
in terms of certain dual fields transforming under a gauge group 
of possibly
different rank, $SU( N_c'(i_0))$, where 
$ N_c'(i_0) = N_f(i_0) - N_c(i_0)$.
The deformed theory is written in terms of these dual 
variables, called 
\emph{dual quarks}, and  certain composites of the 
original fields, called \emph{mesons}. The couplings are then 
restored to  generic
values, thus obtaining a dual description of the original 
theory, with a possibly different gauge group.
Let us start by recalling the rules for obtaining Seiberg 
dual theories starting from a given 
quiver gauge theory \cite{fhhk}. As mentioned above, 
in obtaining a
Seiberg dual theory, one singles out a node $i_0$ of the quiver
as the dualising node.  Seiberg duality alters the 
rank of the gauge
group associated to this node, as well as the quarks in the 
theory. First, one
makes certain combinatorial operations on the quiver ignoring the
superpotential. After obtaining the dual quiver, the
consistency of the dual theory is checked by verifying that
the fields in this theory satisfy F-flatness conditions of a
cubic superpotential. Apart from changing the rank of the gauge
group $SU(N_c(i_0))$, these operations modifies the adjacency 
matrix $C$ of the quiver,  whose entries $C_{ij}$ denote 
the number of arrows from the $i$-th to the $j$-th node. 
Let us point out that to start with we can
write the adjacency matrix as one with all entries
positive, counting only the arrows emanating from the nodes. 
Then the quiver of the Seiberg dual theory can be obtained by applying
the following rules.

Let $i_0$ denote the dualising node. Let the set of nodes
from which the incoming arrows on $i_0$ originate be denoted
$I_{in}$ and the set of nodes on which outgoing arrows from
$i_0$ terminate be denoted $I_{out}$. The Seiberg dual 
quiver is obtained by 
\begin{itemize} 
\item changing the rank $N_c(i_0)$ of the gauge group factor
on the node $i_0$ to
$N'_c(i_0)=N_f(i_0)-N_c(i_0)$, with
$\displaystyle N_f(i_0) = \sum_{j\in I_{in}}C_{j{i_0}}N_c(j)$.
The new node is interpreted as an anti-brane of the original
one, and  
\item deriving the adjacency matrix $\tilde{C}$ of the dual quiver 
from that of the original quiver as,
\begin{equation}
\label{rules}
\dual{C}_{ij} = \begin{cases} C_{ji}, & \text{if} \quad
\text{either}~i_0=i \text{~or~} i_0 =j, \\
C_{ij} - C_{ji_0} C_{i_0i}, & \text{if}\quad i\in I_{out}
\text{~\small\&~}  j\in I_{in}, \\
C_{ij}, & \text{otherwise}.
\end{cases}
\end{equation} 
\end{itemize}
A negative entry in the dual adjacency matrix is drawn as a
reversed arrow on the dual quiver.
We shall see that the entries of $\tilde{C}_{ij}$ count
the morphisms between complexes in $\End{T}$ in our examples.
\section{Quiver theories from the second del Pezzo surface}
\label{delpez2}
\begin{figure}[h]
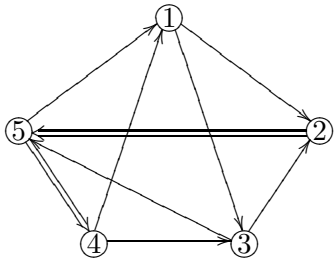

\hbox{\parbox[h]{.6\textwidth}{%
\begin{center}
\mbox{}\hskip 2cm
\xy 
(10,0)*+{4}*\cir{}="D",
(30,0)*+{3}*\cir{}="C",
(0,15)*+{5}*\cir{}="E",
(40,15)*+{2}*\cir{}="B",
(20,30)*+{1}*\cir{}="A"
\ar@{->} "A";"B" <1pt>
\ar@{->} "A";"C" <1pt>
\ar@2{->} "B";"E" <1pt>
\ar@{->} "C";"B" <1pt>
\ar@{->} "C";"E" <1pt>
\ar@{->} "D";"C" <1pt>
\ar@{->} "D";"A" <1pt>
\ar@{->} "E";"A" <1pt>
\ar@2{->} "E";"D" <1pt>
\endxy 
\end{center}
}%
\parbox{.3\textwidth}{
\caption{Quiver $Q$ corresponding to the second del Pezzo
surface}
\label{qdp2}
}
}
\end{figure}

In this section we find out examples of Seiberg dual
theories by explicitly constructing tilting complexes and
working out their endomorphism algebras. In particular, we
consider the quiver theories associated to the complex
cone over the second del Pezzo surface, which arise in a
certain partial resolution of $\C^3/\Z_3\times\Z_3$.
Another partial resolution of $\C^3/\Z_3\times\Z_3$ yields
the complex cone over the zeroth Hirzebruch surface, for which
toric duality predicts two toric phases and hence two toric
quiver theories \cite{fhhu}. That these two theories are
derived equivalent has also been established \cite{braun}.
The quivers in this example have a simple pattern.
There are four nodes in the quivers and one of the 
quivers have the feature that two arrows originate from each
node to go to the adjacent node while two arrows from the
other adjacent node terminate on it. Thus, taking any of
the four nodes as the dualising node leads to the same dual
quiver up to permutations of nodes.
This does not hold in the present case. The quivers
associated to the complex cone over the second del Pezzo
surface have five nodes each. Again, one of them is rather
symmetric in the sense that two arrows emanate from each
node and two arrows terminate on each. But they do not
connect only the adjacent nodes, nor the pair of emanating
or terminating arrows connect to the same nodes. This is the
quiver, shown in Figure~\ref{qdp2}, that we begin with.
The other quivers, as we find, do not have this
feature. 
We then take each node of the quiver in Figure~\ref{qdp2}
in turn as the dualising node and write down the
adjacency matrix of the dual quiver using the rules
\eq{rules}. We then present a tilting complex for 
each case whose endomorphism algebra coincides with the path
algebra of the dual quiver. The homomorphisms between the
direct summands of the tilting complex in each case have to
satisfy certain relations, namely the F-flatness
conditions from the corresponding superpotential. We
write down the dual superpotential in each case and verify that
the homomorphisms we find do satisfy the F-flatness
conditions of the dual superpotential.
However, for the dual theory obtained upon dualising on the
first node, we find that it is easier to find the dual
superpotential by imposing the relations between the
homomorphisms.
Let us start by writing down the quiver data and the gauge
theory data corresponding to the
quiver $Q$ in  Figure~\ref{qdp2}, which we  sometimes
refer to as  \emph{the original quiver}.
The adjacency matrix is 
\begin{equation}
\label{adj:orig}
\begin{split}
C = \begin{pmatrix}
0&1&1&0&0\\
0&0&0&0&2\\
0&1&0&0&1\\
1&0&1&0&0\\
1&0&0&2&0
\end{pmatrix}
\end{split} 
\end{equation} The gauge group is taken to be $\prod_{i=1}^5 SU(n_i)$, that
is $N_c(i)=n_i$ for
$i=1,\cdots , 5$, fixing the representation of the quiver. 
The requirement that the gauge theory is non-anomalous 
constraints the numbers $n_i$. Cancellation of
anomaly requires
\begin{equation}
\sum_j (C - C^{\mathrm{T}})_{ij}\cdot N_c(j) = 0. 
\end{equation} 
Using the adjacency matrix \eq{adj:orig} this yields two
relations among the five $n$'s, namely,
\begin{equation} 
n_1+n_4=n_2+n_5,\qquad n_2+n_3=n_4+n_5.
\end{equation} 
The superpotential of the
$\mathsf{N}=1$ gauge theory corresponding to the original
quiver can be obtained 
as the superpotential of a partially resolved theory from the
superpotential of the $\mathsf{N}=1$ gauge theory on 
$\C^3/\Z_3\times\Z_3$.
The superpotential is  \cite{fhhu}
\begin{equation} 
\label{pot:orig}
\begin{split}
{\cal W} = y_{52}x_{23}x_{34}x_{45} &+ x_{14}y_{45} x_{52}x_{21} +
x_{53}x_{31}x_{15} \\ &- y_{45}x_{53} x_{34} - 
x_{52}x_{23}x_{31}x_{14}x_{45} -
x_{15}y_{52}x_{21}.
\end{split}
\end{equation} 
Each term in the superpotential  is 
a trace over the combinations of chiral fields.
We shall not mention the trace explicitly in this note. 
In \eq{pot:orig} each $x_{ij}$ or $y_{ij}$ denote a
bosonic component of the chiral fields of the $\mathsf{N}=1$ theory.
The subscripts indicate the nodes they connect, as
mentioned in the last section. 

We now proceed to Seiberg dualise this quiver node by node. Let us
note at the outset that since there are five nodes in the
quiver we expect at most five dual theories, including
the one corresponding to the original quiver. Now, as we
Seiberg dualise the five nodes, we obtain five dual
theories, apart from the original one. We then expect one of
these five new theories to coincide with the original one.
Indeed, dualising on the first node we obtain a quiver which
gives back the original quiver after re-labelling the nodes.
Let us start with this case.
\subsection{Dualising on the first node}
\begin{figure}[h]
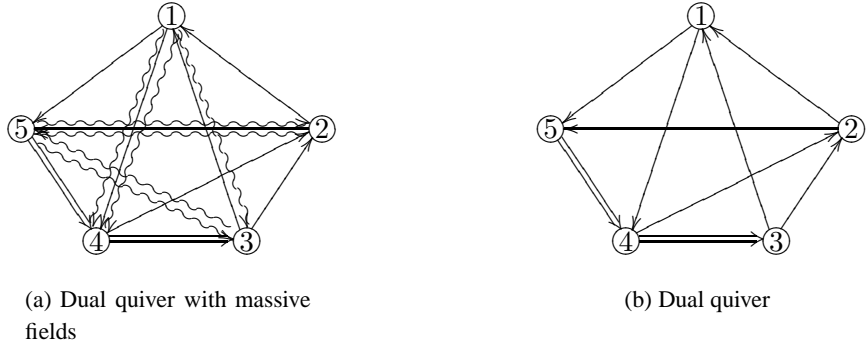

\begin{center}
\subfigure[Dual quiver with massive fields]{%
\label{qdp2:meson1}
\xy
(10,0)*+{4}*\cir{}="D",
(30,0)*+{3}*\cir{}="C",
(0,15)*+{5}*\cir{}="E",
(40,15)*+{2}*\cir{}="B",
(20,30)*+{1}*\cir{}="A"
\ar@{~>} "A";"C" <2pt>
\ar@{->} "A";"D" <0pt>
\ar@{~>} "A";"D" <3pt>
\ar@{~>} "A";"D" <-3pt>
\ar@{->} "A";"E" <1pt>
\ar@{->} "B";"A" <1pt>
\ar@{->} "B";"E" <0pt>
\ar@{~>} "B";"E" <2pt>
\ar@{->} "C";"A" <1pt>
\ar@{->} "C";"B" <1pt>
\ar@{~>} "C";"E" <-2pt>
\ar@2{->} "D";"C" <1pt>
\ar@{->} "D";"B" <1pt>
\ar@2{->} "E";"D" <1pt>
\ar@{~>} "E";"B" <2pt>
\ar@{~>} "E";"C" <-2pt>
\endxy
}%
\hskip 1in
\subfigure[Dual quiver]{%
\label{qdp2:dual1}
\xy
(10,0)*+{4}*\cir{}="D",
(30,0)*+{3}*\cir{}="C",
(0,15)*+{5}*\cir{}="E",
(40,15)*+{2}*\cir{}="B",
(20,30)*+{1}*\cir{}="A"
\ar@{->} "B";"A" <1pt>
\ar@{->} "C";"B" <1pt>
\ar@2{->} "D";"C" <1pt>
\ar@2{->} "E";"D" <1pt>
\ar@{->} "A";"E" <1pt>
\ar@{->} "C";"A" <1pt>
\ar@{->} "D";"B" <1pt>
\ar@{->} "A";"D" <1pt>
\ar@{->} "B";"E" <0pt>
\endxy
}%
\end{center}
\caption{Quivers from Seiberg dualising on node $1$}
\end{figure}

We begin by taking the first node as the dualising node, that is,
$i_0 =1$. Then we have $I_{in} = \{5,4\}$ and $I_{out} =
\{2,3\}$. Now
applying the rules \eq{rules} on the node $1$ we can find
out the adjacency matrix of the dual quiver and its
representation. Obviously, only the rank of the gauge group
associated to this node may change. 
The adjacency matrix of the new quiver turns out to be
\begin{equation}
\begin{split}
\tilde{C} =
\begin{pmatrix}
0&0&0&1&1&\\
1&0&0&-1&1\\
1&1&0&-1&0\\
0&0&1&0&0\\
0&0&0&2&0
\end{pmatrix}.
\end{split} 
\end{equation}
The corresponding quiver is shown in Figure~\ref{qdp2:dual1}.
Let us recall that a negative value of $C_{ij}$ is
represented by an arrow from the $j$-th to the $i$-th node.
Our goal is to re-derive this quiver from the endomorphism
algebra of a tilting complex. For this let us start with the
projectives $P_i$, $i=1,\cdots , 5$, corresponding to the
five nodes and consider the complex $T = \oplus_{i=1}^5 T_i$
in the derived category $D(Q)$ of the original quiver $Q$,
with  direct summands 
\begin{equation}
\begin{split}
T_1:&\quad\xymatrix{0 \ar[r] & \underline{P_4\oplus P_5}
\ar[r]^(.6){\hbox{$\left(\begin{smallmatrix} x_{14}\\ 
-x_{15}\end{smallmatrix}\right)$}} & P_1\ar[r] & 0}\\ 
T_2:&\quad \xymatrix{0\ar[r] & \underline{P_2}\ar[r] & 0}\\
T_3:&\quad \xymatrix{0\ar[r] & \underline{P_3}\ar[r] & 0}\\
T_4:&\quad \xymatrix{0\ar[r] & \underline{P_4}\ar[r] & 0}\\
T_5:&\quad \xymatrix{0\ar[r] & \underline{P_5}\ar[r] & 0}.
\end{split} 
\end{equation}
The zeroth position in the complexes are underlined.
It can be checked that $T$ is a tilting complex,
$\Ext{T_i,T_j}=0$ for all $i,j=1, \cdots , 5$.
We then evaluate the non-zero homomorphisms between each
pair of $T_i$ in the derived category $D(Q)$, thus finding the 
endomorphism algebra of $T$. The result is summarised in
Table~\ref{table:node1}. The entries which are absent from
the table are zero maps in the corresponding homotopy
category. 
\begin{table}[h]
\begin{tabular}{lccccc}
& $T_1$ & $T_2$ & $T_3$ & $T_4$ & $T_5$\\
\hline\hline 
$T_1$ \vline & 
$\left(\left(\begin{smallmatrix}1&0\\0&1\end{smallmatrix}\right),1\right)$
& &
$X_{31}:= \left(\left(\begin{smallmatrix}x_{34}\\0
\end{smallmatrix}\right), 0\right)$ &
$\begin{matrix}
X_{41}:= \left(\left(\begin{smallmatrix}1\\0
\end{smallmatrix}\right), 0\right) \\
Y_{41}:= \left(\left(\begin{smallmatrix} 0 \\x_{45}
\end{smallmatrix}\right), 0\right) \\
Z_{41}:= \left(\left(\begin{smallmatrix} 0\\ y_{45}
\end{smallmatrix}\right), 0\right)
\end{matrix}$ & 
$X_{51}:= \left(\left(\begin{smallmatrix}0\\1
\end{smallmatrix}\right), 0\right)$  
\\			
$T_2$ \vline &
$X_{12}:= \left(\left(\begin{smallmatrix} y_{45} x_{52}\\y_{52}
\end{smallmatrix}\right), 0\right)$ 
& $(1,0)$ && & $
\begin{matrix}
X_{52}:= (x_{52},0) \\
Y_{52}:= (y_{52},0)^{\star}
\end{matrix}$
\\			
$T_3$ \vline &
$X_{13}:= \left(\left(\begin{smallmatrix} x_{45}x_{52}x_{23}\\x_{53}
\end{smallmatrix}\right), 0\right)$ 
& $X_{23}:=(x_{23},0)$ & $(1,0)$ & & $X_{53}:= (x_{53}, 0)^{\star}$
\\			
$T_4$ \vline & & $X_{24}:= (x_{21}x_{14}, 0)$ &
$\begin{matrix}
X_{34}:= (x_{34},0) \\
Y_{34}:= (x_{31}x_{14},0)
\end{matrix}$
& $(1,0)$ &
\\			
$T_5$ \vline && $X_{25}:= (x_{21}x_{15}, 0)^{\star}$ &
$X_{35}:= (x_{31}x_{15}, 0)^{\star}$ &
$\begin{matrix}
X_{45}:= (x_{45},0) \\
Y_{45}:= (y_{45},0)
\end{matrix}$
& $(1,0)$
\end{tabular}
\caption{$\End{T}$ for dualising on node 1. Massive fields marked 
$\star$.}
\label{table:node1}
\end{table}
The homomorphisms listed in the table are the dual fields in
the gauge theory
--- dual quarks and dual mesons --- and the dual superpotential 
has to be  expressed in terms of these fields.
Drawing arrows corresponding to the entries in the table we
obtain the quiver Figure~\ref{qdp2:meson1}.
It can be checked from the table that $X_{31} = X_{34} X_{41}$, 
$Y_{41}=X_{45}X_{51}$ and 
$Z_{41}=Y_{45}X_{51}$. Thus these three homomorphisms are
determined by the others and the corresponding fields are
absent from the superpotential.  Moreover, from
the superpotential $\cal W$ in \eq{pot:orig} we note that 
$x_{31}x_{15}x_{53} = X_{35}X_{53}$
and $y_{52} x_{21} x_{15} = Y_{52} X_{25}$, thus giving mass 
to the four dual fields. In the table they are marked with an
asterisk $(\star)$. These seven fields are to be dropped out
from the quiver and shown as wiggly lines in
Figure~\ref{qdp2:meson1}. Leaving out these fields 
the resulting quiver is as shown in Figure~\ref{qdp2:dual1}, the dual
quiver that we obtained using the rules \eq{rules} above.

Let us briefly digress to discuss the combinatorics of the above
calculations with an example. From the rules 
\eq{rules}, we have
$\tilde{C}_{ij} = C_{ji}$ when $i=1$ or $j=1$. The
effect of this is to invert the arrows on the node 1. In
writing the tilting complex $T$ this has been achieved by
shifting the projective $P_1$ in the summand $T_1$ by one
degree. Indeed, a one-degree shift is the operation in the
derived category that takes a brane to its anti-brane
\cite{douglas}. Furthermore, we have, for example,
$\tilde{C}_{25} = C_{25} - C_{51}C_{12} = 2-1$ from
\eq{rules}, while $\tilde{C}_{52} = C_{52} = 0$. This shows
that one of the two arrows from node $2$ to node $5$ is
``killed" by an arrow which is a composite of arrows from
node $5$ to node $1$ and from $1$ to $2$. The ``killer" and
the ``victim" are represented by the wiggly lines in
Figure~\ref{qdp2:meson1}. Both of them are eliminated in
the dual quiver (none of them may be around at large, after
all!). From Table~\ref{table:node1} it can
be checked that the fields corresponding to the
homomorphisms $Y_{52}$ and $X_{25}$, the latter being the
same composite as stated above, become massive together. 
Similar combinatorics apply to the evaluation of 
$\End{T}$ in all the cases in this note.

We still need to check whether the morphisms listed
in the table satisfy the relations derived from F-flatness
conditions of the dual
superpotential. For this we first need to write the dual
potential by expressing $\cal W$ in terms of the dual fields 
from the table and then adding to it the ``mesonic" part
\cite{beasley,fhhu}. Finally we integrate out the fields
which gain superpotential masses to derive the
dual superpotential.
Thus, let us  first rewrite the superpotential $\cal W$ in 
terms of the
dual fields. To this end, let us first note that the dual 
quarks are $X_{12}$, $X_{13}$, $X_{41}$ and $X_{51}$, that 
is the inverted arrows on the node 1. The mesons are
the composites, $X_{24}$, $X_{25}$, $Y_{34}$ and $X_{35}$. 
The superpotential $\cal W$ expressed in terms of the dual 
fields from Table~\ref{table:node1} assumes the form 
\begin{equation}
\begin{split}
{\cal W}' = Y_{52} X_{23} X_{34} X_{45} &+ Y_{45} X_{52} X_{24} 
+ X_{53} X_{35} \\
& - Y_{45} X_{53} X_{34} - X_{52} X_{23} Y_{34} X_{45} -
Y_{52} X_{25}.
\end{split} 
\end{equation} 
Let us point out that none of the fields in the above
expression correspond to an arrow which either originates or
terminates on the dualising node $1$. That is, the loops
corresponding to the terms do not go through the first node.
It is clear from the expression for ${\cal W}'$ that the 
fields $X_{53}$, $X_{35}$, $Y_{52}$
and $X_{25}$ are massive, as indicated in Table~\ref{table:node1}.
Now, we write the new part of the superpotential involving 
the mesons. The
particular combinations appearing in the superpotential and 
the signs
of the terms in this part can be fixed by noting that 
\label{thumbrules}
\begin{itemize}
\item each field occurs twice and only twice in the
superpotential, 
\item two terms involving any given field appear with
opposite signs,
\item any pair of terms may have only a single field in
common.
\end{itemize}
Using these rules, we get the ``mesonic" part of the superpotential 
\begin{equation}
{\cal W}_m = 
  X_{13} Y_{34} X_{41} + X_{12} X_{25} X_{51}
- X_{12} X_{24} X_{41} - Y_{13} X_{35} X_{51},
\end{equation} 
in which each term corresponds to a loop going through the
\emph{new} first node.
We then integrate out the massive fields from the new superpotential 
${\cal W}' + {\cal W}_m$ by using their equations of motion, for
example,
\begin{equation}
X_{25} = X_{23} X_{34} X_{45},
\end{equation} 
obtained from $\pa\tilde{\cal W} / \pa Y_{52} =0$ etc. Finally 
we obtain the dual
superpotential 
\begin{equation}
\begin{split}
\tilde{\cal W} = X_{12} X_{23} X_{34} X_{45} X_{51} 
&+ Y_{45} X_{52} X_{24} +
X_{13} Y_{34} X_{41} \\
&- Y_{45} X_{51} X_{13} X_{34} - X_{51} X_{23} Y_{34} X_{45} 
- X_{12} X_{24} X_{41}.
\end{split}
\end{equation} 
It can now be verified, by using the expressions in 
Table~\ref{table:node1} and
the F-flatness conditions of the original superpotential, 
that the dual fields
satisfy the eleven F-term equations coming from $\tilde{\cal W}$. 
Let us discuss the two non-trivial cases that need be
treated specially.

The expression $\pa\tilde{\cal W}/\pa X_{13} =  
Y_{34} X_{41} - X_{34} Y_{45}
X_{51}$, in terms of the original fields is 
\begin{equation}
\begin{pmatrix}
x_{31} x_{14}\\ - x_{34} y_{45}
\end{pmatrix} 
= x_{31}
\begin{pmatrix}
x_{14} \\ - y_{15}
\end{pmatrix},
\end{equation} 
where we used the F-term equation, 
$\pa {\cal W} / \pa x_{53} = x_{31} x_{15} - x_{34}
y_{45}=0$. While not zero as it
is, this expression is homotopic to zero
with homotopy $(\cdots , 0, x_{31}, 0, \cdots )$,
 as can be seen from the following diagram 
\begin{equation}
\begin{split}
\xymatrix{
0 \ar[r] & \underline{P_4\oplus P_5}
\ar[r]^(.6){\hbox{$\left(\begin{smallmatrix} x_{14}\\ 
-x_{15}\end{smallmatrix}\right)$}} 
\ar[d]_{\hbox{$\left(\begin{smallmatrix}x_{31}x_{14}\\
-x_{31}x_{15}\end{smallmatrix}\right)$}}
& P_1\ar[r]\ar@{-->}[dl]|{\hbox{$x_{31}$}} & 0  \\
0\ar[r] & \underline{P_3}\ar[r] & 0}
\end{split} 
\end{equation} 
Thus, $\pa\tilde{\cal W}/\pa X_{13}=0$ in the homotopy category 
of the complexes
$\{T_i, i=1,\cdots , 5\}$ and hence in the resulting derived 
category.

The expression $\pa\tilde{\cal W} / \pa X_{12} = 
X_{23} X_{34} X_{45} X_{51} -
X_{24} X_{41}$ can similarly be shown to be zero in 
the derived category, using
the relation $\pa {\cal W}/\pa x_{52} = x_{23} x_{34} x_{45} 
- x_{21} x_{15}=0$  
and the following diagram 
\begin{equation}
\begin{split}
\xymatrix{
0 \ar[r] & \underline{P_4\oplus P_5}
\ar[r]^(.6){\hbox{$\left(\begin{smallmatrix} x_{14}\\ 
-x_{15}\end{smallmatrix}\right)$}} 
\ar[d]_{\hbox{$\left(\begin{smallmatrix}x_{21}x_{14}\\
-x_{21}x_{15}\end{smallmatrix}\right)$}}
& P_1\ar[r]\ar@{-->}[dl]|{\hbox{$x_{21}$}} & 0  \\
0\ar[r] & \underline{P_2}\ar[r] & 0}
\end{split} 
\end{equation} 

As mentioned earlier, the quiver Figure~\ref{qdp2:dual1} can be
mapped to the original quiver Figure~\ref{qdp2} by the 
following re-labelling of the nodes: 
\entrymodifiers={+<1pc>[o][F-]}
\[\xymatrix{ 1 \ar[r] & 3\ar[r] & 4 \ar[r] & 5 \ar[r] 
& 2 \ar[r] & 1 
}\]
\entrymodifiers={}
\subsection{Dualising on the second node}
\begin{figure}[h]
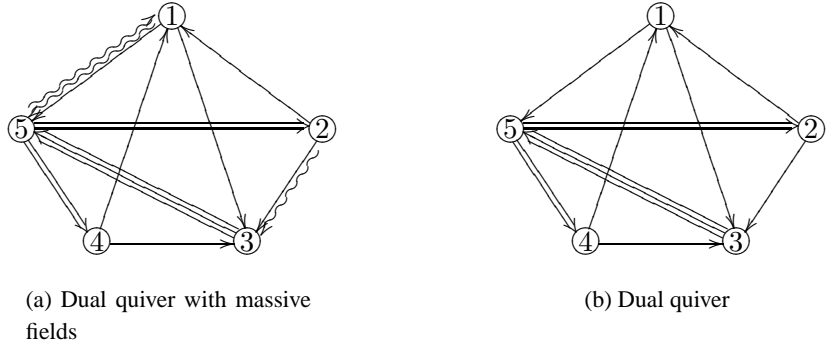

\begin{center}
\subfigure[Dual quiver with massive fields]{%
\label{qdp2:meson2}
\xy
(10,0)*+{4}*\cir{}="D",
(30,0)*+{3}*\cir{}="C",
(0,15)*+{5}*\cir{}="E",
(40,15)*+{2}*\cir{}="B",
(20,30)*+{1}*\cir{}="A"
\ar@{->} "A";"C" <1pt>
\ar@{->} "A";"E" <0pt>
\ar@{~>} "A";"E" <-2pt>
\ar@{->} "B";"A" <1pt>
\ar@{->} "B";"C" <1pt>
\ar@{~>} "B";"C" <3pt>
\ar@3{->} "C";"E" <-1pt>
\ar@{->} "D";"A" <0pt>
\ar@{->} "D";"C" <-1pt>
\ar@{~>} "E";"A" <4pt>
\ar@2{->} "E";"B" <1pt>
\ar@2{->} "E";"D" <-1pt>
\endxy
}%
\hskip 2cm 
\subfigure[Dual quiver]{%
\label{qdp2:dual2}
\xy
(10,0)*+{4}*\cir{}="D",
(30,0)*+{3}*\cir{}="C",
(0,15)*+{5}*\cir{}="E",
(40,15)*+{2}*\cir{}="B",
(20,30)*+{1}*\cir{}="A"
\ar@{->} "A";"C" <1pt>
\ar@{->} "A";"E" <0pt>
\ar@{->} "B";"A" <1pt>
\ar@{->} "B";"C" <1pt>
\ar@3{->} "C";"E" <-1pt>
\ar@{->} "D";"A" <0pt>
\ar@{->} "D";"C" <-1pt>
\ar@2{->} "E";"B" <1pt>
\ar@2{->} "E";"D" <-1pt>
\endxy
}%
\end{center}
\caption{Quivers from Seiberg dualising on node $2$}
\end{figure}

Let us now perform Seiberg duality on the same quiver $Q$ in 
Figure~\ref{qdp2} taking the node $2$ as the dualising node.
Again, applying the rules \eq{rules} to calculate the
adjacency matrix of the dual quiver we obtain 
\begin{equation}
\tilde{C} = \begin{pmatrix}
0&0&1&0&0\\
1&0&1&0&0\\
0&0&0&0&1\\
1&0&1&0&0\\
-1&2&-2&2&0
\end{pmatrix}. 
\end{equation} 
The resulting quiver with the massive fields is shown
in Figure~\ref{qdp2:meson2}, while the dual quiver is
shown in Figure~\ref{qdp2:dual2}. The details of the
calculation is similar to the previous case and we shall not
discuss them here.
We reproduce the path algebra of the quiver
Figure~\ref{qdp2:dual2} as the endomorphism algebra of the
tilting complex $T = \oplus_{i=1}^5 T_i$, with the direct
summands 
\begin{equation}
\begin{split}
T_1:&\quad \xymatrix{0\ar[r] & \underline{P_1}\ar[r] & 0}\\
T_2:&\quad\xymatrix{0 \ar[r] & \underline{P_1\oplus P_3} 
\ar[r]^(.6){\hbox{$\left(\begin{smallmatrix}
x_{21}\\ -x_{23}\end{smallmatrix}\right)$}} &
P_2\ar[r] & 0}\\ 
T_3:&\quad \xymatrix{0\ar[r] & \underline{P_3}\ar[r] & 0}\\
T_4:&\quad \xymatrix{0\ar[r] & \underline{P_4}\ar[r] & 0}\\
T_5:&\quad \xymatrix{0\ar[r] & \underline{P_5}\ar[r] & 0}
\end{split} 
\end{equation} 
The morphisms in $\End{T}$ are shown in Table~\ref{table:node2}.
\begin{table}[h]
\begin{tabular}{lccccc}
& $T_1$ & $T_2$ & $T_3$ & $T_4$ & $T_5$ \\
\hline\hline 
$T_1$ \vline & $(1,0)$ &
 & $X_{31}:= (x_{31},0) $ & & 
$\begin{matrix}
X_{51}:= (x_{52}x_{21}, 0) \\
Y_{51}:= (y_{52}x_{21}, 0)^{\star}
\end{matrix}$  \\  		
$T_2$ \vline & $X_{12}:= 
\left(\left(\begin{smallmatrix} 1\\0\end{smallmatrix}\right), 0\right)$ & 
$\left(\left(\begin{smallmatrix}1&0\\0&1\end{smallmatrix}\right),1\right)
$ &
$\begin{matrix}
X_{32}:= 
\left(\left(\begin{smallmatrix} x_{31} \\0\end{smallmatrix}\right),
0\right)\\
Y_{32}:= 
\left(\left(\begin{smallmatrix} 0\\1\end{smallmatrix}\right),0\right)
\end{matrix}$ 
&& 
$X_{52}:= \left(\left(\begin{smallmatrix} 0\\x_{53}\end{smallmatrix}
\right), 0\right)$ 
 \\   
$T_3$ \vline & & & $(1,0)$ & & 
$\begin{matrix}
X_{53}:= (x_{53}, 0) \\
Y_{53}:= (x_{52}x_{23} , 0) \\
Z_{53}:= (y_{52}x_{23}, 0)
\end{matrix}$  \\   
$T_4$ \vline & $X_{14}:= (x_{14}, 0) $ & & $X_{34}:= (x_{34}, 0)$ 
& $(1,0)$ & \\ 
$T_5$ \vline & $X_{15}:= (x_{15}, 0)^{\star}$ &
$\begin{matrix}
X_{25}:=
\left(\left(\begin{smallmatrix} x_{15}\\x_{34}x_{45}
\end{smallmatrix}\right),0\right) \\
Y_{25}:= 
\left(\left(\begin{smallmatrix} x_{14}y_{45} \\ x_{31}x_{14}x_{45}
\end{smallmatrix}\right),0\right) \\
\end{matrix}$  &&
$\begin{matrix}
X_{45}:= (x_{45}, 0) \\
Y_{45}:= (y_{45}, 0)
\end{matrix}$ 
& $(1,0)$
\end{tabular}
\caption{$\End{T}$ for dualising on node 2. 
Massive fields marked $\star$.}
\label{table:node2}
\end{table}
From the table, we see that $X_{32} = X_{31} X_{12}$ and 
$X_{52}=X_{53} Y_{32}$, and are therefore dropped from the
quiver as well as from the dual superpotential.
Moreover, from the superpotential $\cal W$, the 
term $x_{15} y_{52} x_{21} =
Y_{51} X_{15}$. This makes the two dual quarks massive.
These are marked with an asterisk in the above table.

The calculation of the dual superpotential proceeds 
as in the previous case. 
In this case, the dual quarks are $X_{12}$, $Y_{32}$, $X_{25}$ 
and $Y_{25}$,
while the mesons are $X_{51}$ $Y_{51}$, $Y_{53}$ and $Z_{53}$.
As in the previous case, we derive the dual superpotential 
for the present example to be 
\begin{equation}
\begin{split}
\tilde{\cal W}=
Z_{53} X_{34} X_{45} &+ X_{14} Y_{45} X_{51} + 
X_{25} X_{53} X_{31} X_{12} +
Y_{25} Y_{53} Y_{32} \\
& - Y_{45} X_{53} X_{34} - Y_{53} X_{31} X_{14} X_{45} 
- Y_{25} X_{51} X_{12} -
X_{25} Z_{53} Y_{32}.
\end{split} 
\end{equation} 
The  relations ensuing
from the F-flatness conditions of the superpotential
$\tilde{\cal W}$ are satisfied by the morphisms in
Table~\ref{table:node2} upon using the F-flatness conditions 
of the original superpotential $\cal W$. Two of the cases are
non-trivial. These  are 
\begin{equation}
\pa\tilde{\cal W} / \pa X_{25} = 
\begin{pmatrix}
y_{52} x_{21} \\ - y_{52} x_{23}
\end{pmatrix}
\quad \txt{and} \quad
\pa\tilde{\cal W} / \pa Y_{25} = 
\begin{pmatrix}
x_{52} x_{21} \\ - x_{52} x_{23}
\end{pmatrix},
\end{equation} 
which can be seen to be zero in the derived category from the 
following diagrams
respectively
\begin{equation}
\begin{split}
\xymatrix{
0 \ar[r] & \underline{P_1\oplus P_3}
\ar[r]^(.6){\hbox{$\left(\begin{smallmatrix} x_{21}\\ 
-x_{23}\end{smallmatrix}\right)$}} 
\ar[d]_{\hbox{$\left(\begin{smallmatrix}y_{52}x_{21}\\
-y_{52}x_{23}\end{smallmatrix}\right)$}}
& P_2\ar[r]\ar@{-->}[dl]|{\hbox{$y_{52}$}} & 0  \\
0\ar[r] & \underline{P_5}\ar[r] & 0}
\qquad\txt{and}\qquad
\xymatrix{
0 \ar[r] & \underline{P_1\oplus P_3}
\ar[r]^(.6){\hbox{$\left(\begin{smallmatrix} x_{21}\\ 
-x_{23}\end{smallmatrix}\right)$}} 
\ar[d]_{\hbox{$\left(\begin{smallmatrix}x_{52}x_{21}\\
-x_{52}x_{23}\end{smallmatrix}\right)$}}
& P_2\ar[r]\ar@{-->}[dl]|{\hbox{$x_{52}$}} & 0  \\
0\ar[r] & \underline{P_5}\ar[r] & 0}
\end{split} 
\end{equation} 
\subsection{Dualising on the third node}
\begin{figure}[h]
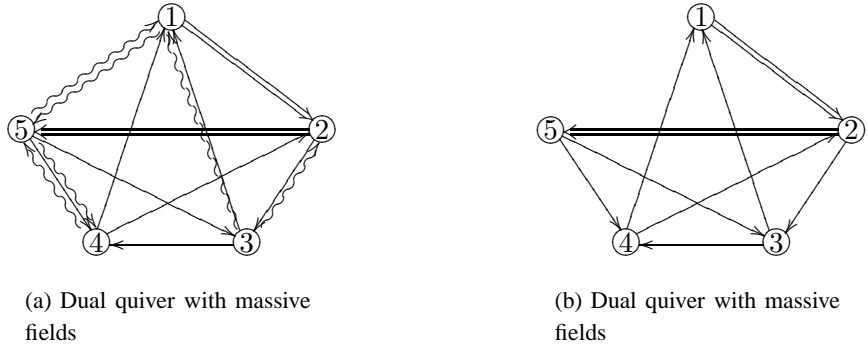

\begin{center}
\subfigure[Dual quiver with massive fields]{%
\label{qdp2:meson3}
\xy
(10,0)*+{4}*\cir{}="D",
(30,0)*+{3}*\cir{}="C",
(0,15)*+{5}*\cir{}="E",
(40,15)*+{2}*\cir{}="B",
(20,30)*+{1}*\cir{}="A"
\ar@2{->} "A";"B" <1pt>
\ar@{~>} "A";"E" <1pt>
\ar@{->} "B";"C" <0pt>
\ar@2{->} "B";"E" <1pt>
\ar@{->} "C";"A" <1pt>
\ar@{~>} "C";"A" <3pt>
\ar@{~>} "C";"B" <-2pt>
\ar@{->} "C";"D" <1pt>
\ar@{->} "D";"A" <1pt>
\ar@{->} "D";"B" <1pt>
\ar@{~>} "D";"E" <2pt>
\ar@{~>} "E";"A" <2pt>
\ar@{->} "E";"D" <1pt>
\ar@{~>} "E";"D" <3pt>
\ar@{->} "E";"C" <0pt>
\endxy
}%
\hskip 1in
\subfigure[Dual quiver with massive fields]{%
\label{qdp2:dual3}
\xy
(10,0)*+{4}*\cir{}="D",
(30,0)*+{3}*\cir{}="C",
(0,15)*+{5}*\cir{}="E",
(40,15)*+{2}*\cir{}="B",
(20,30)*+{1}*\cir{}="A"
\ar@2{->} "A";"B" <1pt>
\ar@{->} "B";"C" <1pt>
\ar@2{->} "B";"E" <1pt>
\ar@{->} "C";"A" <1pt>
\ar@{->} "C";"D" <1pt>
\ar@{->} "D";"A" <1pt>
\ar@{->} "D";"B" <1pt>
\ar@{->} "E";"D" <1pt>
\ar@{->} "E";"C" <0pt>
\endxy
}%
\end{center}
\caption{Quivers from Seiberg dualising on node $3$}
\label{qdp2:dual22}
\end{figure}

We repeat the exercise of performing Seiberg duality with now the
third node as the dualising node. The calculations are
similar to the previous cases and we only present the
results. 
The adjacency matrix of the dual quiver obtained by applying
the rules \eq{rules} is 
\begin{equation}
{\tilde{C}} = \begin{pmatrix}
0&1&0&0&0\\
-1&0&1&-1&2\\
1&0&0&1&0\\
1&0&0&0&0&\\
0&0&1&1&0
\end{pmatrix}. 
\end{equation} 
The dual quivers, with and without the massive
fields, are shown in Figure~\ref{qdp2:dual22} as before.
The tilting complex $T$ in the derived category $D(Q)$ of
the original quiver is taken as $T=\oplus_{i=1}^5T_i$ with
the direct summands
\begin{equation}
\begin{split}
T_1:&\quad \xymatrix{0\ar[r] & \underline{P_1}\ar[r] & 0}\\
T_2:&\quad \xymatrix{0\ar[r] & \underline{P_2}\ar[r] & 0}\\
T_3:&\quad\xymatrix{0 \ar[r] & \underline{P_1\oplus P_4}
\ar[r]^(.6){\hbox{$\left(\begin{smallmatrix}
x_{31}\\ -x_{34}\end{smallmatrix}\right)$}} &
P_3\ar[r] & 0} \\
T_4:&\quad \xymatrix{0\ar[r] & \underline{P_4}\ar[r] & 0}\\
T_5:&\quad \xymatrix{0\ar[r] & \underline{P_5}\ar[r] & 0}.
\end{split} 
\end{equation} 
The morphisms in the endomorphism algebra $\End{T}$ are
tabulated in Table~\ref{table:node3}.
\begin{table}[h]
\begin{tabular}{lccccc}
& $T_1$ & $T_2$ & $T_3$ & $T_4$ & $T_5$ \\
\hline\hline 
$T_1$ \vline &$(1,0)$ &
$\begin{matrix}
X_{21}:= (x_{21}, 0) \\
Y_{21}:= (x_{23}x_{31},0) 
\end{matrix}$  
&  & & $X_{51}:= (x_{53}x_{31},0)^{\star}$
\\                  
$T_2$ \vline & & $(1,0)$ &
$X_{32}:=\left(\left(
\begin{smallmatrix}
 x_{14}x_{45}x_{52}\\ x_{45}y_{52}
\end{smallmatrix}
\right),0\right)
$ &&
$\begin{matrix}
X_{52}:= (x_{52}, 0) \\
Y_{52}:= (y_{52}, 0) 
\end{matrix}$  
\\                 
$T_3$ \vline &
$\begin{matrix}
X_{13}:= 
\left(\left(\begin{smallmatrix} 1\\0\end{smallmatrix}\right),0\right)
\\Y_{13}:= 
\left(\left(\begin{smallmatrix} 0\\x_{14}\end{smallmatrix}\right),0\right) 
\end{matrix}$  
&
$ X_{23}:=
\left(\left(\begin{smallmatrix} x_{21}\\0\end{smallmatrix}\right),0\right) 
$ 
&
$\left(\left(\begin{smallmatrix}1&0\\0&1\end{smallmatrix}\right),1\right)$ 
&
$X_{43}:= 
\left(\left(\begin{smallmatrix} 0\\1\end{smallmatrix}\right),0\right) $
&
\\		
$T_4$ \vline 
& $X_{14}:=(x_{14},0)$ 
& $X_{24}:=(x_{23}x_{34},0)$
& & $(1,0)$ 
&$X_{54}:= (x_{53}x_{34},0)^{\star}$
\\		
$T_5$ \vline & $X_{15}:= (x_{15},0)^{\star}$
& &
$
X_{35}:= 
\left(\left(\begin{smallmatrix} x_{15}\\y_{45}\end{smallmatrix}\right),0\right)
$  
&
$\begin{matrix}
X_{45}:= (x_{45}, 0) \\
Y_{45}:= (y_{45}, 0)^{\star} 
\end{matrix}$  
& $(1,0)$
\end{tabular}
\caption{$\End{T}$ for dualising on node 3. Massive fields marked $\star$.}
\label{table:node3}
\end{table}
Here again we find that the two fields 
$X_{23} = X_{21}X_{13}$ and $Y_{13}=X_{14}X_{43}$ are
determined by others and hence dropped.
Moreover, from the terms $x_{53}x_{31}x_{15} =
X_{51}X_{15}$ and $y_{45}x_{53}x_{34} = Y_{45}X_{54}$ in the
superpotential $\cal W$, we see that four dual
quarks become massive.

The dual superpotential is evaluated as before. In this case
the dual quarks are $X_{13}$, $X_{43}$, $X_{32}$ and $X_{35}$, while the mesons
are $X_{51}$, $X_{24}$, $Y_{21}$ and $X_{54}$.
In the present case the dual superpotential takes the form 
\begin{equation} 
\begin{split}
\tilde{\cal W} = Y_{52} X_{24} X_{45} &+ X_{14} X_{43} X_{35} X_{52} X_{21} 
+ X_{32} Y_{21} X_{13} \\
&- X_{52} Y_{21} X_{14} X_{45} - X_{35} Y_{52} X_{21} X_{13} - X_{32} X_{24}
X_{43}.
\end{split}
\end{equation} 
The corresponding relations from the F-flatness conditions
of $\tilde{\cal W}$ are satisfied by the morphisms in Table~\ref{table:node3}
upon using the F-flatness conditions ensuing from $\cal W$. Again there are two
non-trivial relations, namely, 
\begin{equation} 
\pa\tilde{\cal W} / \pa X_{35} = 
\begin{pmatrix} -x_{53} x_{31} \\ x_{53} x_{34}
\end{pmatrix}
\quad\txt{and}\quad 
\pa\tilde{\cal W} / \pa X_{21} = 
\begin{pmatrix} x_{23} x_{31} \\ -x_{23} x_{34}
\end{pmatrix},
\end{equation} 
which can be shown to be zero in the derived category by considering the
following diagrams respectively.
\begin{equation}
\begin{split}
\xymatrix{
0 \ar[r] & \underline{P_1\oplus P_4}
\ar[r]^(.6){\hbox{$\left(\begin{smallmatrix} x_{31}\\ 
-x_{34}\end{smallmatrix}\right)$}} 
\ar[d]_{\hbox{$\left(
\begin{smallmatrix}-x_{53} x_{31}\\x_{53} x_{34}
\end{smallmatrix}\right)$}}
& P_3\ar[r]\ar@{-->}[dl]|{\hbox{$-x_{53}$}} & 0  \\
0\ar[r] & \underline{P_5}\ar[r] & 0}
\qquad\txt{and}\qquad
\xymatrix{
0 \ar[r] & \underline{P_1\oplus P_4}
\ar[r]^(.6){\hbox{$\left(\begin{smallmatrix} x_{31}\\ 
-x_{34}\end{smallmatrix}\right)$}} 
\ar[d]_{\hbox{$\left(\begin{smallmatrix}x_{23} x_{31}\\-x_{23} x_{34}
\end{smallmatrix}\right)$}}
& P_3\ar[r]\ar@{-->}[dl]|{\hbox{$x_{23}$}} & 0  \\
0\ar[r] & \underline{P_2}\ar[r] & 0}
\end{split} 
\end{equation} 
\subsection{Dualising on the fourth node}
\begin{figure}[h]
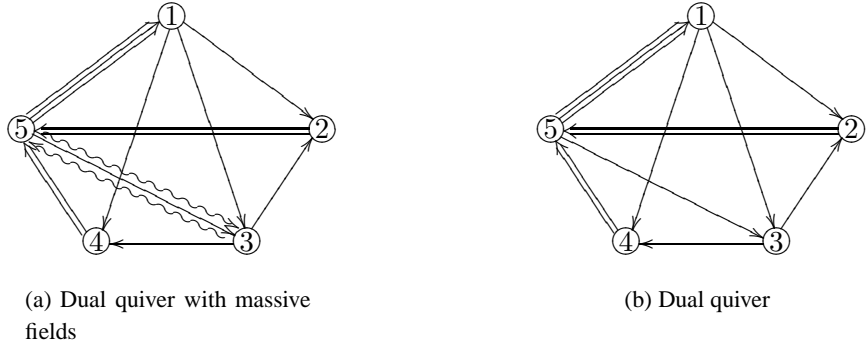

\begin{center}
\subfigure[Dual quiver with massive fields]{
\label{qdp2:meson4}
\xy 
(10,0)*+{4}*\cir{}="D",
(30,0)*+{3}*\cir{}="C",
(0,15)*+{5}*\cir{}="E",
(40,15)*+{2}*\cir{}="B",
(20,30)*+{1}*\cir{}="A"
\ar@{->} "A";"B" <1pt>
\ar@{->} "A";"C" <1pt>
\ar@{->} "A";"D" <1pt>
\ar@2{->} "B";"E" <1pt>
\ar@{->} "C";"B" <1pt>
\ar@{->} "C";"D" <1pt>
\ar@2{->} "D";"E" <2pt>
\ar@3{->} "E";"A" <1pt>
\ar@{~>} "E";"C" <3pt>
\ar@{->} "E";"C" <0pt>
\ar@{~>} "C";"E" <3pt> 
\endxy
}
\hskip 1in
\subfigure[Dual quiver]{
\label{qdp2:dual4}
\xy 
(10,0)*+{4}*\cir{}="D",
(30,0)*+{3}*\cir{}="C",
(0,15)*+{5}*\cir{}="E",
(40,15)*+{2}*\cir{}="B",
(20,30)*+{1}*\cir{}="A"
\ar@{->} "A";"B" <1pt>
\ar@{->} "A";"C" <1pt>
\ar@{->} "A";"D" <1pt>
\ar@2{->} "B";"E" <1pt>
\ar@{->} "C";"B" <1pt>
\ar@{->} "C";"D" <1pt>
\ar@2{->} "D";"E" <1pt>
\ar@3{->} "E";"A" <1pt>
\ar@{->} "E";"C" <-1pt>
\endxy 
}
\end{center}
\caption{Quivers from Seiberg dualising on node 4}
\end{figure}

Next, by  choosing the fourth node for dualisation we obtain the
adjacency matrix of the dual quiver 
\begin{equation} 
\tilde{C} = \begin{pmatrix}
0&1&1&1&-2\\
0&0&0&0&2\\
0&1&0&1&-1\\
0&0&0&0&2\\
1&0&0&0&0
\end{pmatrix}.
\end{equation} 
by applying the rules \eq{rules}. This furnishes the dual
quiver shown in Figure~\ref{qdp2:dual4}. The quiver with
massive fields is shown in Figure~\ref{qdp2:meson4}.
The gauge theory corresponding to the dual quiver
Figure~\ref{qdp2:dual4} is a toric one found earlier through
toric duality \cite{fhhu}.
The path algebra of the quiver Figure~\ref{qdp2:dual4} 
coincides with the endomorphism algebra of the 
tilting complex $T=\oplus_{i=1}^5 T_i$ in $D(Q)$, with 
direct summands
\begin{equation}
\begin{split}
T_1:&\quad \xymatrix{0\ar[r] & \underline{P_1}\ar[r] & 0}\\
T_2:&\quad \xymatrix{0\ar[r] & \underline{P_2}\ar[r] & 0}\\
T_3:&\quad \xymatrix{0\ar[r] & \underline{P_3}\ar[r] & 0}\\
T_4:&\quad \xymatrix{0 \ar[r] & \underline{P_5\oplus P_5}
\ar[r]^(.6){\hbox{$\left(\begin{smallmatrix}
x_{45}\\ -y_{45}\end{smallmatrix}\right)$}} & P_4\ar[r] & 0}\\
T_5:&\quad \xymatrix{0\ar[r] & \underline{P_5}\ar[r] & 0}.
\end{split} 
\end{equation} 
The morphisms in $\End{T}$ are shown in 
Table~\ref{table:node4}.
We have the fields $X_{14} = X_{15} Y_{54}$ and $Y_{14} = X_{15} X_{54}$
determined in terms of others  and hence dropped.
Moreover, in  the original superpotential $\cal W$, the term
$y_{45} x_{53} x_{34} = X_{53} Y_{35}$, which makes the two dual quarks massive.
\begin{table}[h]
\begin{tabular}{lccccc}
& $T_1$ & $T_2$ & $T_3$ & $T_4$ & $T_5$ \\	
\hline\hline
$T_1$ \vline & $\left(\begin{smallmatrix}
1\\0\end{smallmatrix}\right)$ & $X_{21}:=(x_{21},0)$ & $X_{31}:= (x_{31}, 0)$ & 
$X_{41}:= \left(\left(
\begin{smallmatrix} x_{52}x_{23}x_{31}\\x_{52}x_{21} \end{smallmatrix}
\right),0\right) $
& \\ 						
$T_2$ \vline & & $(1,0)$ & && $\begin{matrix}X_{52}:= (x_{52},0) \\ Y_{52}:=
(y_{52},0)\end{matrix}$ \\ 			
$T_3$ \vline & &$X_{23}:= (x_{23},0)$ & $(1,0)$ &
$X_{43}:= \left(\left(
\begin{smallmatrix} y_{52}x_{23}\\x_{53}\end{smallmatrix}\right),0\right)$ & 
$X_{53}:= (x_{53}, 0)^{\star}$ \\			
$T_4$ \vline & 
$\begin{matrix} X_{14}:= \left(\left(\begin{smallmatrix}
x_{15}\\0\end{smallmatrix}\right), 0\right)\\
Y_{14}:= \left(\left(\begin{smallmatrix}0\\x_{15}\end{smallmatrix}\right),
0\right)\end{matrix} $ &
&& $\left(\left(\begin{smallmatrix}1&0\\0&1\end{smallmatrix}\right),1\right)$ &
$\begin{matrix}
X_{54}:= \left(\left(\begin{smallmatrix}0\\1\end{smallmatrix}\right),0\right) \\
Y_{54}:= \left(\left(\begin{smallmatrix}1\\0\end{smallmatrix}\right),0\right)
\end{matrix}$ \\ 				
$T_5$ \vline  & $\begin{matrix}
X_{15}:= (x_{15},0)\\
Y_{15}:= (x_{14}x_{45},0)\\
Z_{15}:= (x_{14}y_{45},0)
\end{matrix}$ &&
$\begin{matrix}
X_{35}:= (x_{34}x_{45},0)\\
Y_{35}:= (x_{34}y_{45},0)^{\star}
\end{matrix}
$
&& $(1,0)$
\end{tabular}
\caption{$\End{T}$ for dualising on node 4. Massive fields marked $\star$.}
\label{table:node4}
\end{table}
In this case
the surviving dual quarks are $X_{54}$, $Y_{54}$, 
$X_{41}$ and $X_{43}$, while the mesons
are $Y_{15}$, $Z_{15}$, $X_{35}$ and $Y_{35}$.
The dual superpotential is now given by 
\begin{equation}
\begin{split}
\tilde{\cal W} = X_{43}X_{35}Y_{54} &+ X_{15}Y_{52}X_{21} +X_{41}Z_{15}X_{54}
 - X_{35}Y_{52}X_{23} \\
& -X_{41}Y_{15}Y_{54}
- X_{15}X_{54}X_{43}X_{31}
- Z_{15}X_{52}X_{21}
-X_{31}Y_{15}X_{52}X_{23}
\end{split} 
\end{equation} 
Again, the maps from Table~\ref{table:node4} satisfy the corresponding
F-flatness conditions. The non-trivial ones are
\begin{equation}
\pa\tilde{\cal W} / \pa X_{41} = 
\begin{pmatrix}
x_{14} x_{45} \\ -x_{14} y_{45}
\end{pmatrix}
\quad \txt{and} \quad
\pa\tilde{\cal W} / \pa X_{43} = 
\begin{pmatrix}
x_{34} x_{45} \\ -x_{34} y_{45}
\end{pmatrix}
\end{equation} 
which can be seen to be zero in the derived category from the following diagrams
respectively
\begin{equation}
\begin{split}
\xymatrix{
0 \ar[r] & \underline{P_5\oplus P_5}
\ar[r]^(.6){\hbox{$\left(\begin{smallmatrix} x_{45}\\-y_{45}
\end{smallmatrix}\right)$}} 
\ar[d]_{\hbox{$\left(\begin{smallmatrix} x_{14} x_{45} \\ 
-x_{14} y_{45}
\end{smallmatrix}\right)$}}
& P_4\ar[r]\ar@{-->}[dl]|{\hbox{$x_{14}$}} & 0  \\
0\ar[r] & \underline{P_1}\ar[r] & 0}
\qquad\txt{and}\qquad
\xymatrix{
0 \ar[r] & \underline{P_5\oplus P_5}
\ar[r]^(.6){\hbox{$\left(\begin{smallmatrix} x_{45}\\
-y_{45}\end{smallmatrix}\right)$}} 
\ar[d]_{\hbox{$\left(\begin{smallmatrix} x_{34} x_{45} \\ 
-x_{34} y_{45}
\end{smallmatrix}\right)$}}
& P_4\ar[r]\ar@{-->}[dl]|{\hbox{$x_{34}$}} & 0  \\
0\ar[r] & \underline{P_3}\ar[r] & 0}
\end{split} 
\end{equation} 
\subsection{Dualising on the fifth node}
\begin{figure}[h]
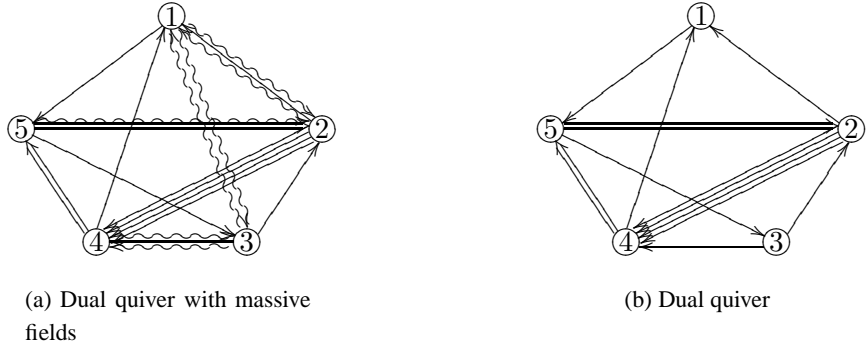

\begin{center}
\subfigure[Dual quiver with massive fields]{%
\label{qdp2:meson5}
\xy
(10,0)*+{4}*\cir{}="D",
(30,0)*+{3}*\cir{}="C",
(0,15)*+{5}*\cir{}="E",
(40,15)*+{2}*\cir{}="B",
(20,30)*+{1}*\cir{}="A"
\ar@{~>} "A";"B" <2pt>
\ar@{~>} "A";"C" <2pt>
\ar@{->} "A";"E" <1pt>
\ar@{->} "B";"A" <0pt>
\ar@{~>} "B";"A" <2pt>
\ar@{->} "B";"D" <-3pt>
\ar@{->} "B";"D" <-1pt>
\ar@{->} "B";"D" <1pt>
\ar@{->} "B";"D" <3pt>
\ar@{~>} "C";"A" <2pt>
\ar@{->} "C";"B" <-2pt>
\ar@{->} "C";"D" <0pt>
\ar@{~>} "C";"D" <2pt>
\ar@{->} "D";"A" <1pt>
\ar@{~>} "D";"C" <2pt>
\ar@2{->} "D";"E" <1pt>
\ar@2{->} "E";"B" <1pt>
\ar@{~>} "E";"B" <3pt>
\ar@{->} "E";"C" <0pt>
\endxy
}%
\hskip 1in
\subfigure[Dual quiver]{%
\label{qdp2:dual5}
\xy
(10,0)*+{4}*\cir{}="D",
(30,0)*+{3}*\cir{}="C",
(0,15)*+{5}*\cir{}="E",
(40,15)*+{2}*\cir{}="B",
(20,30)*+{1}*\cir{}="A"
\ar@{->} "A";"E" <1pt>
\ar@{->} "B";"A" <1pt>
\ar@{->} "B";"D" <-3pt>
\ar@{->} "B";"D" <-1pt>
\ar@{->} "B";"D" <1pt>
\ar@{->} "B";"D" <3pt>
\ar@{->} "C";"B" <-2pt>
\ar@{->} "C";"D" <2pt>
\ar@{->} "D";"A" <1pt>
\ar@2{->} "D";"E" <1pt>
\ar@2{->} "E";"B" <1pt>
\ar@{->} "E";"C" <0pt>
\endxy
}%
\end{center}
\caption{Quivers from Seiberg dualising on node 5}
\end{figure}

Finally, we  dualise the quiver $Q$ in Figure~\ref{qdp2} by
performing Seiberg duality on the fifth node. Again, we
apply the rules \eq{rules} to derive the adjacency matrix of
the dual quiver as 
\begin{equation}
\tilde{C}=
\begin{pmatrix}
0&-1&0&0&1\\
0&0&0&0&0\\
0&1&0&0&0\\
1&-4&-1&0&2\\
0&2&1&0&0
\end{pmatrix}.
\end{equation} 
The corresponding dual quiver is shown in
Figure~\ref{qdp2:dual5}. The one with massive fields is shown
in Figure~\ref{qdp2:meson5}.
The path algebra of the dual quiver Figure~\ref{qdp2:dual5} is
obtained as  the endomorphism algebra of the tilting complex
$T=\oplus_{i=1}^5 T_i$ in the derived category $D(Q)$, with
direct summand
\begin{equation}
\begin{split}
T_1:&\quad \xymatrix{0\ar[r] & \underline{P_1}\ar[r] & 0}\\
T_2:&\quad \xymatrix{0\ar[r] & \underline{P_2}\ar[r] & 0}\\
T_3:&\quad \xymatrix{0\ar[r] & \underline{P_3}\ar[r] & 0}\\
T_4:&\quad \xymatrix{0\ar[r] & \underline{P_4}\ar[r] & 0}\\
T_5:&\quad\xymatrix{0 \ar[r] & \underline{P_2\oplus P_2\oplus P_3}
\ar[r]^(.7){\hbox{$\left(\begin{smallmatrix}
x_{52}\\ y_{52}\\ x_{53}\end{smallmatrix}\right)$}} &
P_5\ar[r] & 0}. 
\end{split}
\end{equation}
The homomorphisms in $\End{T}$ are shown in
Table~\ref{table:node5}.
\begin{table}[h]
\begin{tabular}{lccccc}
& $T_1$ & $T_2$ & $T_3$ & $T_4$ & $T_5$ \\
\hline\hline 		
$T_1$ \vline & $(1,0)$ & $X_{21}:=(x_{21},0)^{\star}$ & 
$X_{31}:=(x_{31},0)^{\star}$ & & 
$X_{51}:= \left(\left(\begin{smallmatrix}
0\\x_{21}\\-x_{31}
\end{smallmatrix}\right),0\right)$ 
\\   			
$T_2$ \vline & 
$\begin{matrix}
X_{12}:= (x_{15}x_{52},0)\\
Y_{12}:= (x_{15}y_{52},0)^{\star}
\end{matrix}$
& $(1,0)$ & & 
$\begin{matrix}
X_{42}:= (x_{45}x_{52},0)\\
Y_{42}:= (x_{45}y_{52},0) \\
Z_{42}:= (y_{45}x_{52},0)\\
W_{42}:= (y_{45}y_{52},0)
\end{matrix}$
&  \\			
$T_3$ \vline &
$X_{13}:= (x_{15}x_{53}, 0)^{\star}$ &
$X_{23}:= (x_{23}, 0)$ &
$(1,0)$ &
$\begin{matrix}
X_{43}:= (x_{45}x_{53}, 0)\\
Y_{43}:= (y_{45}x_{53}, 0)^{\star}
\end{matrix}$
\\ 			
$T_4$ \vline &
$X_{14}:= (x_{14}, 0)$ & &
$X_{34}:= (x_{34}, 0)^{\star}$ & $(1,0)$ &
$\begin{matrix}
X_{54}:=
\left(\left(\begin{smallmatrix}x_{21}x_{14}\\0\\-x_{34}
\end{smallmatrix}\right),
0\right) \\
Y_{54}:= \left(\left(\begin{smallmatrix}x_{23}x_{31}x_{14}\\
-x_{23}x_{34}\\
0\end{smallmatrix}\right), 0\right)
\end{matrix}$
\\			
$T_5$ \vline &
&
$\begin{matrix}
X_{25}:= \left(\left(\begin{smallmatrix}1\\0\\0\end{smallmatrix}\right),
0\right)\\
Y_{25}:= \left(\left(\begin{smallmatrix}0\\1\\0\end{smallmatrix}\right), 
0\right)\\
Z_{25}:= \left(\left(\begin{smallmatrix}0\\0\\x_{23}\end{smallmatrix}\right),
0\right)
\end{matrix}$
&
$X_{35}:= \left(\left(\begin{smallmatrix}0\\0\\1\end{smallmatrix}\right),
0\right)$
& &
$\left(\left(\begin{smallmatrix}1&0&0\\0&1&0\\0&0&1
\end{smallmatrix}\right),1\right)$ 
\end{tabular}
\caption{$\End{T}$ for dualising on node 5. Massive fields marked $\star$.}
\label{table:node5}
\end{table}
From the table we note that $Z_{25} = X_{23}X_{35}$ and is thus dropped.
Moreover, from the superpotential $\cal W$, the terms 
$y_{45}x_{53}x_{34} = Y_{43}X_{34}$, $x_{15}x_{53}x_{21} = X_{13}X_{31}$ and
$x_{15}y_{52}x_{21} = Y_{12}X_{21}$, thus giving masses to the dual quarks.

The calculation of the dual superpotential for this case
presents some interesting subtleties. 
The rules mentioned in 
page~\pageref{thumbrules} can not be used to fix the
dual superpotential. In order to  evaluate
the dual superpotential systematically, we first write down 
all possible terms in the superpotential and then retain
the ones which will be compatible with the relations between
the fields in Table~\ref{table:node5}. Indeed, we use
only a few relations to determine the superpotential
completely. The other relations can then be checked to
be satisfied, thus ensuring the consistency of the procedure.

Let us discuss some details of this
calculation in brief. 
The dual quarks in the gauge theory, as can be read off from
Table~\ref{table:node5}, are $X_{25}$, $Y_{25}$, $X_{35}$, 
$X_{51}$, $X_{54}$ and $Y_{54}$
while the mesons are $X_{12}$, $Y_{12}$, $X_{13}$, $X_{42}$, 
$Y_{42}$, $Z_{42}$, $W_{42}$, $X_{43}$ and $Y_{43}$.
First, we express the superpotential $\cal W$ in terms of
the dual fields from the table, as usual. It assumes 
the following form 
\begin{equation}
\begin{split}
{\cal W}'= Y_{42}X_{23}X_{34} &+ X_{14}Z_{42}X_{21} +
X_{31}X_{13} \\ 
&- Y_{43}X_{34} - X_{42}X_{23}X_{31}X_{14} - Y_{12} X_{21}.
\end{split} 
\end{equation} 
Clearly, $X_{34}$, $Y_{43}$, $X_{31}$ and $X_{13}$ have
superpotential mass. Before writing down the mesonic
part let us note that the massive fields
$X_{34}$, $X_{21}$ and $X_{31}$ are not in the above list of dual
quarks and mesons. Hence they will not appear in 
the mesonic part ${\cal W}_m$ of the superpotential.
The variations of the dual superpotential with respect to these 
fields, therefore, have no other possible contribution than
those arising from the variations of ${\cal W}'$. 
The variations of ${\cal W}'$ with respect to $X_{34}$,
$X_{21}$ and $X_{31}$  yield  
\begin{equation}
\label{threefld}
Y_{43} = Y_{42} X_{23}, \quad  Y_{12} = X_{14} Z_{42},\quad
X_{13} = X_{14}X_{42}X_{23},
\end{equation} 
respectively.
Substituting these in the expression of ${\cal W}'$ we find
that the superpotential ${\cal W}'$ vanishes. This means
that the dual superpotential is expressed solely in terms of
the perturbations of the original superpotential. This
feature is unique to the case in hand among the examples
studied in this note.

The dual superpotential thus consists only of the
mesonic part. There are twenty five terms in the mesonic
part, consisting
of combinations of the dual quarks and mesons listed above.
In writing the mesonic, alias dual, superpotential
we have to decide which of these twenty five terms will be
retained as part of  ${\cal W}_m$ and also determine
their signs. It can be checked
that the rules mentioned in page~\pageref{thumbrules} fail
in this case to determine all the terms. In order
to find out this part of the superpotential we first write
down all the twenty five terms with both the possible signs
for each. The resulting potential takes the following form.
\begin{equation}
\label{twn5}
\begin{split}
\hat{W} = \pm X_{54} X_{43} X_{35} & \pm
Y_{54}X_{43} X_{35}
\pm  X_{54} Y_{43} X_{35} \pm Y_{54} Y_{43} X_{35} \\
&\pm \underline{X_{51} X_{12} X_{25}} \pm X_{51} X_{12} Y_{25} 
\pm X_{51} Y_{12} X_{25} \pm \underline{X_{51} Y_{12} Y_{25}}
\pm \underline{X_{51} X_{13} X_{35}} \\
&\pm X_{54} X_{42} X_{25} \pm X_{54} X_{42} Y_{25} 
\pm Y_{54} X_{42} X_{25} \pm Y_{54} X_{42} Y_{25} \\
&\pm X_{54} Y_{42} X_{25} \pm X_{54} Y_{42} Y_{25} 
\pm Y_{54} Y_{42} X_{25} \pm Y_{54} Y_{42} Y_{25} \\
&\pm X_{54} Z_{42} X_{25} \pm X_{54} Z_{42} Y_{25} 
\pm Y_{54} Z_{42} X_{25} \pm Y_{54} Z_{42} Y_{25} \\
&\pm X_{54} W_{42} X_{25} \pm X_{54} W_{42} Y_{25} 
\pm Y_{54} W_{42} X_{25} \pm Y_{54} W_{42} Y_{25}.
\end{split} 
\end{equation} 
We then consider variations of this new potential 
with respect to some of the fields. From
Table~\ref{table:node5} we
find out which of the terms may  be retained in order for
these variations to vanish. Let us illustrate this procedure
with two examples.

There are only two terms in ${\cal W}'$ which contain
$X_{43}$, namely the first two in \eq{twn5}. 
Variation of $\hat{\cal W}$ with respect to $X_{43}$ should
yield the F-flatness
conditions to be satisfied by the fields in 
Table~\ref{table:node5}. From the table we find that
$X_{35} X_{54} = -x_{34}$, while $X_{35} Y_{54}=0$. Hence only
the second one can be retained. The sign is not yet fixed,
however. Next, let us consider the terms containing 
$X_{51}$ in $\hat{W}$. There are
five such terms, written in the second line in \eq{twn5}.
Varying with respect to $X_{51}$  we find that the
variation can be made equal to 
\begin{equation}
\psi:=
x_{15} \cdot\begin{pmatrix} 
x_{52} \\y_{52} \\ x_{53}
\end{pmatrix}
\end{equation} 
if we retain only the three terms underlined with same sign
for all the three. Here we used the
F-flatness conditions of the superpotential $\cal W$.
This morphism is zero in the homotopy category of the
complexes $T_i$, as can be seen from the diagram 
\begin{equation} 
\xymatrix{0 \ar[r] & \underline{P_2\oplus P_2\oplus P_3}
\ar[r]^(.7){\hbox{$\left(\begin{smallmatrix}
x_{52}\\ y_{52}\\
x_{53}\end{smallmatrix}\right)$}}\ar[d]_{\hbox{$\psi$}} &
P_5\ar[r]\ar@{-->}[dl]|{\hbox{$x_{15}$}} & 0\\
0 \ar[r] & \underline{P_1}\ar[r] & 0} 
\end{equation} 

Repeating this exercise considering variations of $\hat{W}$ with
respect to $X_{54}$ and $Y_{54}$, we can single out the
terms which appear in the dual superpotential, not
completely fixed yet though. The potential $\hat{W}$ after
dropping out the extra terms from \eq{twn5} takes the form 
\begin{equation}
\begin{split}
\hat{W} = &\alpha (X_{51} X_{12} X_{25} + X_{51} X_{14}
X_{42} X_{23} X_{35} + X_{51} X_{14} Z_{42} Y_{25})  \\
&\beta (X_{54} Z_{42} X_{25} + X_{54} W_{42} Y_{25} + X_{54}
Y_{42} X_{23} X_{35}) \\
&\gamma ( Y_{54} X_{42} X_{25} + Y_{54} Y_{42} Y_{25} +
Y_{54} X_{43} X_{35}),
\end{split}
\end{equation} 
where we have used the relations \eq{threefld} to substitute
the expressions for the fields determined in terms of others.
Here $\alpha$, $\beta$, $\gamma$ are signs ($\pm 1$) to be
determined. To this end, we consider the variation of
$\hat{W}$ with respect to $X_{35}$. Using the
expressions  of the dual fields from Table~\ref{table:node5}  
we write the variation in terms of the original fields to obtain
\begin{equation}
\pa \hat{W} /\pa X_{35} = 
\begin{pmatrix}
\beta\cdot x_{21} x_{14} x_{45} y_{52} x_{23}+\gamma\cdot x_{23}
x_{31} x_{14} x_{45} x_{53}\\
\alpha\cdot x_{21} x_{14} x_{45} x_{52} x_{23} - \gamma\cdot
x_{23} x_{34} x_{45} x_{53}\\
-\alpha\cdot x_{31} x_{14} x_{45} x_{52} x_{23} - \beta\cdot
x_{34} x_{45} y_{52} x_{23}
\end{pmatrix}.
\end{equation}
Now, using the F-flatness conditions ensuing from the
variations of $\cal W$,
we can set this variation to zero by choosing $\alpha =1$,
$\beta=-1$ and $\gamma =1$. This fixes all the signs 
in $\hat{W}$, which yields the mesonic superpotential 
${\cal W}_m$, which in turn is the dual superpotential
$\tilde{\cal W}=0$. That is, finally we have obtained the
dual superpotential
\begin{equation}
\begin{split}
\tilde{\cal W} = X_{51} X_{12} X_{25} &+ X_{51} X_{14}
X_{42} X_{23} X_{35} + X_{51} X_{14} Z_{42} Y_{25}  \\
&- X_{54} Z_{42} X_{25} - X_{54} W_{42} Y_{25} - 
X_{54} Y_{42} X_{23} X_{35} \\
&+ Y_{54} X_{42} X_{25} + Y_{54} Y_{42} Y_{25} +
Y_{54} X_{43} X_{35}.
\end{split} 
\end{equation}  
Let us point out that we have used variations of the
potential $\hat{W}$ with respect to only five out of the
fourteen dual fields
to fix the dual superpotential $\tilde{\cal W}$. 
It can now be verified that the variations of the dual
superpotential with respect to all the dual fields vanish.
In other words, the relations arising from the 
F-flatness conditions of the dual theory are satisfied
by the morphisms in Table~\ref{table:node5}.
\section{The general picture}
\label{discus}
From the combinatorics of the 
examples discussed above and  earlier ones
\cite{berdoug,braun} there emerges a general scheme to
obtain Seiberg dual quiver gauge theories with
superpotentials. Let us conclude this note by discussing this
scheme. The idea is to start with a quiver $Q$ along with a
superpotential $\cal W$ of the associated $\mathsf{N}=1$ gauge
theory. We choose one node of the quiver as the
dualising node, denoted $i_0$. We then  write down a tilting
complex $T$ in the bounded derived category $D(Q)$ of the 
path algebra of $Q$ using the incoming arrows on $i_0$. 
The tilting complex $T$
has the form $\oplus_{i=1}^N T_i$, where $N$ equals the
number of nodes in $Q$, where each direct summand $T_i$ 
consists of the projective $P_i$ from the path algebra
$Q\operatorname{-mod}$ in the zeroth degree, if
$i\neq i_0$. The summand $T_{i_0}$ is a two-term complex,
which is of the form 
\begin{equation} 
T_{i_0}:\qquad\xymatrix{ 0 \ar[r] & \hbox{$
\underline{\displaystyle
\oplus_{j\in I_{in}}P_j}$}\ar[r]^(.6){\hbox{$\left(
\goth{p}_{i_0j}\right)$}} & P_{i_0}\ar[r] & 0,
} 
\end{equation} 
where $j\in I_{in}$ and $\goth{p}_{i_0j}$ denote the 
morphisms in $Q\operatorname{-mod}$ corresponding to the arrows 
terminating  on $i_0$ as well as the associated chiral fields. 
By interpreting the  endomorphism algebra of $T$ 
as the path algebra of another quiver, we obtain the dual
quiver $\tilde{Q}$. We now write down the dual
superpotential starting from the original one $\cal W$. 

To this end let us define $\goth{q}_{ki_0}$, with $k\in
I_{out}$ to denote the morphisms corresponding to the arrows
emanating from $i_0$ and the associated fields. The relevant
part of the superpotential of the gauge theory associated to
the quiver $Q$ is 
\begin{equation}
{\cal W} = \goth{p}\cdot {\goth{M}} \cdot\goth{q}, 
\end{equation} 
where, again we refrain from mentioning the trace over gauge
indices explicitly. Here $\goth{M}_{jk}$  is a matrix, in
general, with $j\in I_{out}$ and $k\in I_{in}$ corresponding
to (possibly composite) morphism in $Q\operatorname{-mod}$ 
from the $k$-th to the $j$-th node. The F-flatness
conditions of $\cal W$ imposes 
$\goth{p}\cdot\goth{M} = 0 = \goth{M}\cdot\goth{q}$. 

Similarly, we define dual fields corresponding to the morphisms in
$\tilde{Q}\operatorname{-mod}$, which appear in the dual
superpotential. The dual quarks are denoted by the morphisms 
$\tilde{\goth{p}}$ and $\tilde{\goth{q}}$, in the same
convention as in the last paragraph and each of these 
corresponds to a moiety of $\goth{M}$ as
\begin{equation}
\label{meso1} 
\goth{M} =\tilde{\goth{q}}\cdot\tilde{\goth{p}}. 
\end{equation} 
In order to find the dual fields, we need to solve 
\eq{meso1}.  The solution is not
necessarily unique; the dual fields introduced earlier
correspond to some consistent choice of solutions for this
equation. In fact, this equation does not make sense in the special cases 
in which $\goth{M}$ is a single field, as occurs in considering the McKay 
quiver corresponding to the orbifold $\C^3/\Z_3$. 
Generally, the dual superpotential can be written in
terms of these dual fields by introducing the mesons
\begin{equation}
\label{meso2}
\tilde{\goth{M}}:= \goth{q}\cdot\goth{p}.
\end{equation}
The expressions \eq{meso1} and \eq{meso2} clearly bring out
the duality between the fields we are discussing. 
We write the dual superpotential as a sum of the original
superpotential, written in terms of the dual fields and the
mesonic part, that is the deformation of the original
superpotential, as
\begin{equation}
\label{dsup}
\tilde{\cal W}  = \goth{M}\cdot\tilde{\goth{M}} -
\tilde{\goth{p}}\cdot\tilde{\goth{M}}\cdot\tilde{\goth{q}} ,
\end{equation} 
where in the first term, in re-expressing the original
superpotential $\cal W$ we used cyclic properties of the
trace over gauge indices. Moreover,  $\goth{M}$ must be
expressed in terms of the dual fields in this expression.

Finally, let us verify that the dual fields thus defined
satisfy relations ensuing from the F-flatness conditions of
the dual superpotential \eq{dsup}.
The F-flatness conditions obtained from the variations of
$\tilde{\cal W}$ with respect to the mesons 
$\tilde{\goth{M}}$ are of the form 
$\goth{M} -
\tilde{\goth{q}}\cdot\tilde{\goth{p}} = 0 $, which is 
satisfied by virtue of \eq{meso1}, by construction.
The ones ensuing from the variations with respect to the
dual quarks $\tilde{\goth{p}}$ or $\tilde{\goth{q}}$ involve
$\tilde{\goth{M}}$. These can be shown to be zero in the
homotopy category of the complexes, and hence in the derived
category $D(Q')$, since $\tilde{\goth{M}}$ is a morphism
homotopic to zero as can be shown by considering the diagram 
with $k\in I_{in}$
\begin{equation} 
\xymatrix{ %
0 \ar[r] & \hbox{$\underline{
\oplus_{j\in I_{in}}P_j}$}\ar[r]^(.6){
\hbox{$\goth{p}$}} \ar[d]_{\hbox{$\tilde{\goth{M}}$}} 
& P_{i_0}\ar[r]\ar@{-->}[dl]|{\hbox{$\goth{q}$}} & 0 \\
0 \ar[r] & P_k \ar[r] & 0 &
}%
\end{equation} 
generally, by using \eq{meso2}. This indeed is what happened 
in the examples of the last section.
This furnishes a general method to obtain Seiberg dual
quiver gauge theories, exemplified by the ones in the last
section.
Let us point out that this method does not depend on the
underlying geometry of the vacuum moduli space of the gauge theory.
\section{Summary}
\label{concl}
In this note we have obtained Seiberg dual quiver gauge
theories with $\mathsf{N}=1$ supersymmetry as theories 
associated to quivers whose path algebras are derived
equivalent. The vacuum moduli space of these gauge theories  
are the same, namely, the complex cone over the second del Pezzo
surface. We have established the derived equivalence of the 
path algebras by explicitly constructing tilting complexes 
in the derived category of the path algebras. The endomorphisms of
the tilting complex furnish the dual quarks and mesons of
the dual theory and satisfy the F-flatness conditions
derived from the corresponding dual superpotentials.
In one of the examples we found that it even turns out to be
easier to obtain the dual superpotential from the
endomorphism algebra of the tilting complex.
This furnishes a non-trivial
explicit verification of the conjecture that Seiberg duality is a
derived equivalence of physical gauge theories with 
superpotentials \cite{berdoug}. We constructed the
tilting complexes explicitly for five different cases, by
dualising on each node in turn. As expected, one of the new
quivers thus obtained matches with the original one up to
permutation of nodes. One other matches with the quiver
obtained by toric duality. 

We then verified in each case that
the endomorphism algebra of the resulting tilting complex
yields a Seiberg dual theory. This, in fact, turns out to be a
general pattern in writing the tilting complexes in similar
cases. On the basis of the examples in this note as well as
the ones studied earlier, we presented a general scheme for
obtaining Seiberg dual theories with superpotentials. In
particular, we
showed how to construct the dual quiver from a tilting
complex in the path algebra of the original quiver one
starts with and presented a method to write the dual
superpotential. We showed that the F-flatness conditions
ensuing from the dual superpotential are satisfied by the
morphisms in the endomorphism algebra of the dual quiver in
general.

We have, by this procedure, obtained Seiberg
dual theories, which are not toric. This exemplifies the
fact that derived equivalence is more general than toric
duality.  The rules \eq{rules} obtained earlier, however, 
apply to all these cases and
in fact, the morphisms in $\End{T}$ are precisely counted 
by these rules. This enables one to explore
the complete moduli space of $\mathsf{N}=1$ gauge theories
associated with
representations of a quiver. In this note we obtained at
different phases of the moduli space of gauge theories associated 
with the second del
Pezzo surface which were unreachable by toric duality alone. 
These phases are connected by renormalisation group flows.

While not unexpected on general grounds, 
it would be interesting to explicitly
check if the
general pattern of obtaining Seiberg dual theories
mentioned above work
also for other non-trivial quiver theories.
It would be nice to have a proof of the fact that, although
the general construction does not depend directly on the
underlying geometry of the vacuum moduli space of the gauge
theory, the construction indeed produce gauge theories with
the same moduli space.
As this approach is algebraic not depending 
on the detailed geometry of the underlying space,
it would be interesting to see if this sheds some light on
Seiberg duality in the presence of fluxes, for example,
theories with discrete torsion.
We hope to return to some of these issues in future.
\section*{Acknowledgment}
KR thanks S Sengupta, U Chattopadhyay and especially A King for 
extremely useful discussions and the 
Abdus Salam ICTP for hospitality where part of this work was
done.
The work of SM is supported by NSF grant PHY--0244801.

\end{document}